\documentclass[sigplan,11pt,nonacm=true]{acmart}
\usepackage{blindtext}
\usepackage{listings}
\usepackage{stfloats}
\usepackage{svg}
\usepackage{multirow}
\usepackage{xcolor,colortbl}
\definecolor{light-gray}{gray}{0.95}

\AtBeginDocument{
  \providecommand\BibTeX{{
    \normalfont B\kern-0.5em{\scshape i\kern-0.25em b}\kern-0.8em\TeX}}}

\setcopyright{acmcopyright}
\copyrightyear{2021}
\acmYear{2021}
\acmDOI{10.1145/1122445.1122456}

\acmConference[ICCQ '21]{ICCQ '21: The First International Conference on Code Quality}{March 27, 2021}{Moscow, Russia}
\acmBooktitle{ICCQ '21: The First International Conference on Code Quality, March 27, 2021, Moscow, Russia}
\acmPrice{15.00}
\acmISBN{978-1-4503-XXXX-X/18/06}

\begin{document}

\title{Raising Security Awareness using Cybersecurity Challenges in Embedded Programming Courses}

\author{Tiago Espinha Gasiba}
\email{tiago.gasiba@siemens.com}
\orcid{0000-0003-1462-6701}
\author{Samra Hodzic}
\orcid{0000-0001-8189-9860}
\email{samra.hodzic@siemens.com}
\affiliation{
  \institution{Siemens AG}
  \streetaddress{Otto-Hahn-Ring 6}
  \city{Munich}
  \state{Bavaria}
  \country{Germany}
  \postcode{81379}
}

\author{Ulrike Lechner}
\affiliation{
  \institution{Universität der Bundeswehr München}
  \streetaddress{Werner-Heisenberg-Weg 39}
  \city{Neubiberg}
  \country{Germany}
  \postcode{85579}
}
\email{ulrike.lechner@unibw.de}
\orcid{0000-0002-4286-3184}

\author{Maria Pinto-Albuquerque}
\affiliation{
  \institution{Instituto Universitário de Lisboa (ISCTE-IUL), ISTAR}
  \streetaddress{Av. das Forças Armadas}
  \city{Lisboa}
  \country{Portugal}
  \postcode{1649-026}
}
\email{maria.albuquerque@iscte-iul.pt}
\orcid{0000-0002-2725-7629}

\renewcommand{\shortauthors}{T. E. Gasiba, S. Hodzic, U. Lechner, and M. Pinto-Albuquerque}

\begin{abstract}
Security bugs are errors in code that, when exploited, can lead to serious software vulnerabilities.
These bugs could allow an attacker to take over an application and steal information.
One of the ways to address this issue is by means of awareness training.
The Sifu platform was developed in the industry, for the industry, with the aim to raise software developers' awareness of secure coding.
This paper extends the Sifu platform with three challenges that specifically address embedded programming courses, and describes how to implement these challenges, while also evaluating the usefulness of these challenges to raise security awareness in an academic setting.
Our work presents technical details on the detection mechanisms for software vulnerabilities and gives practical advice on how to implement them.
The evaluation of the challenges is performed through two trial runs with a total of 16 participants.
Our preliminary results show that the challenges are suitable for academia, and can even potentially be included in official teaching curricula.
One major finding is an indicator of the lack of awareness of secure coding by undergraduates.
Finally, we compare our results with previous work done in the industry and extract advice for practitioners.
\end{abstract}



\maketitle

\section{Introduction}

Security vulnerabilities originate in programming errors, that, when left in code, can be exploited by malicious parties and lead to severe security incidents.
Examples of vulnerabilities are, e.g., Shellshock, HeartBleed, POODLE, and DirtyCOW.
The presence of these errors in software code is an indicator of poor code quality \cite{code_quality_ref}.
Consequences of security breaches include, among others, leakage of confidential information, denial of service, and privilege escalation \cite{sec_0}.
Over the last years, the number of vulnerabilities in software has been steadily increasing and the corresponding financial consequences, e.g. for affected companies, is exceeding several billion dollars \cite{Apextechservices2017}.
One reason that might justify this increase in vulnerabilities is the ever increasing complexity of code and systems.

In an industrial context, several methods exist to address code quality, e.g. using Static Application Security Testing (SAST) Tools \cite{oyetoyan2018myths}, IDE plugins, performing threat and risk analysis, and code reviews.
In \cite{code_quality_1}, McIntosh et al. discovered that code review coverage and expertise participation have a significant link with software quality.
While these methods are typically used in the industry, another important aspect should also be considered, namely students from academia, as these will be the next generation workforce.
Furthermore, students might not have access to all of these methods, or might even lack training in cybersecurity.

In this work we look at raising cybersecurity awareness on secure programming for students in an academic setting.
Our work is based on exploring and using CyberSecurity Challenges (CSC) in the academia.
CyberSecurity Challenges is a novel serious game \cite{serious} which aims to raise awareness on secure coding of software developers in the industry \cite{Gasiba2021s}.
These games are well investigated, and are showing very promising results in an industrial setting.
However, these games have, until now, not been explored in an academic setting.

The current research extends previous work \cite{Gasiba2020s} by developing C++ challenges in an industrial secure coding awareness platform (Sifu platform).
Although other providers exist that offer programming courses, the Sifu platform targets especially industrial environments.
The design of the new challenges hereby presented differ from challenges previously implemented in the Sifu platform since they target embedded programming courses. Furthermore, evaluation is done in an academic setting, which allows to compare industry and academia.
In particular, this work compares the perceived benefits of using the awareness platform in academia and industry.
The results of this comparison might differ due to the different background of students and professional software developers.
Therefore we aim to evaluate the suitability of CSC games in the academia, provide details on the implementation of embedded programming challenges, and aid universities to adopt the platform to assist teaching secure programming of embedded systems.
The industry also benefits from the latter, as these students will be better prepared for the industry demands in terms of secure software development.

In the following we introduce three C++ challenges with different security vulnerabilities.
In these challenges, the security vulnerabilities that will be introduced are, according to the authors' experiences in the industry, likely to happen in embedded systems programming.
These are: side-channel vulnerability, invalid memory access, and race condition.
Along with the challenge description, we will address the learning goal of the challenge to raise awareness on secure coding.
We also give details on how the Sifu platform can assess the vulnerability in the code by means of automatically testing that a participant's solution follows secure coding guidelines.
The main contributions of this work are the following:
\begin{itemize}
    \item design of three defensive C++ programming challenges in the industry which aim to raise secure coding  awareness 
    \item base the challenges on security vulnerabilities common in embedded systems
    \item details on how to test that solutions from a participant follow secure coding
    \item preliminary results on the acceptance of the challenges by students in academia
    \item insight into implementation and improvements for practitioners
\end{itemize}

The current work's main target is to use the Sifu platform, which was developed in the industry, extend it, and analyze how it applies in academia.
We aim to answer the following research questions:
\begin{itemize}
    \item[\textbf{RQ1}] How can the Sifu platform be used in academia?
    \item[\textbf{RQ2}] How to decide whether a challenge based on side-channel, invalid memory access, and race condition was correctly solved?
    \item[\textbf{RQ3}] How do students  perceive the challenges presented through the Sifu platform?
\end{itemize}

In \autoref{sec:two}, we present previous work related to our research.
\autoref{sec:three} discusses our approach to design defensive C++ challenges.
We first briefly describe the Sifu platform, which is used to integrate these challenges to be used for training.
The main part of this section is to introduce the implemented challenges and which security vulnerabilities they contain.
We also present our methodology for assessing the presence of the security vulnerabilities in the code.
Additionally, we discuss the evaluation of the developed challenges and platform in the academia.
The results are presented and discussed in \autoref{sec:four}.
Three groups of results are provided: evaluation of the challenge assessment methods, analysis of two test runs including a survey and semi-structured interview, and comparison with our results to previous results for the industrial context.
Finally, \autoref{sec:five} summarizes our work and briefly discusses possible next steps.

\section{Related Work}
\label{sec:two}
Previous research shows that software developers lack secure programming awareness and skills \cite{others_2,Gasiba2020a}. 
Gasiba et al. \cite{Gasiba2020d} introduced CyberSecurity Challenges (CSCs) to raise secure coding awareness of software developers in the industry. CSCs are serious games that refine the popular CTF format and adapt it to the industry. Gasiba et al. \cite{gasiba_re19} researched the constraints and requirements for delivering a cybersecurity challenge which can cover secure coding from an industry perspective.
One important outcome of this research is that the challenges should focus on the defensive perspective, and not on the offensive.
In their work, they introduced a new platform \cite{Gasiba2020s}, which the authors call Sifu. The platform performs an automatic assessment of challenges in terms of compliance to secure coding standards and guidelines. It uses an artificial intelligence coach, which guides the participant throughout the challenge. Their work presents results that indicate that the Sifu platform's CSC events are adequate for raising secure coding awareness of software developers in the industry.
Ruef et al. \cite{ruef2016build} designed a similar training, however their platform is not focused on a specific programming language, and includes offensive challenges.

Tabassum et al. \cite{sec_1} and Whitney et al. \cite{sec_2}, present the importance of secure coding guidelines and standards in the software development life-cycle.
Given the lack of knowledge on secure coding, developers tend to search online resources for answers and solutions.
However, Kurachi et al. \cite{others_6} and Zhang et al \cite{others_5} show that these solutions are not always adequate and blind usage of these solutions can lead to additional problems.

Static analysis is an evolving approach to evaluate programs based exclusively on their source code without running them.
Clang is a C/C++/Objective-C open-source compiler \cite{sca_5}.
It is a continually developing initiative sponsored by large companies such as Apple, Microsoft, and Google.
Clang is currently a popular method for designing new static analyzers.
Its modular architecture is one of Clang's strength \cite{sca_4}.
The Sifu platform makes use of this technology provided by Clang to implement security assessments of challenges.

In \cite{sca_1, sca_2, sca_3,arusoaie2017comparison}, the researchers compare different open-source static analysis tool available for C/C++. 
The authors have developed their C/C++ applications and introduced various vulnerabilities in the application.
They use these applications to check the tool's capabilities to detect the introduced vulnerabilities.
The researchers have also presented a study comparing commercial static code analysis tools for detecting vulnerabilities in a software source code.

In our work, we use the semi-structured interviews methodology as given by Wilson et al. \cite{semi_inter}.
We also use the definition of awareness as given by Hänsch et al. \cite{others_3}.
In their work, they specify awareness as having three components: perception, protection and behavior.
Perception relates to knowledge of threats, protection relates to knowing available mechanisms to protect against these threats, and finally behavior relates to actual individual behavior, e.g. as in actively writing secure code.

\section{Embedded Challenges}
\label{sec:three}

In this work, we address three different cybersecurity challenges, targeting specific security vulnerabilities. This work's challenges are Sorting - Time Side Channel, Complex Factory (invalid memory access), and TOC-TOU Race Condition.
All challenges were implemented in the C++ programming language and integrated into the Sifu platform.
\autoref{sec_guidlines} presents these challenges along with security vulnerabilities and guidelines contained in them. 
In \autoref{sec:sifu_platform}, we briefly introduce the Sifu Platform.
Next, we describe implementation details of the assessment of the cybersecurity vulnerabilities of the three challenges.
We also discuss the evaluation of the implementation of the challenges.
Finally we refer to the setup and results of the empirical study performed with participants from academia.
Notice that the proposed vulnerability detection methods run additionally to several already existing vulnerability detection mechanisms in the Sifu platform.

\begin{table*}[http]
 \caption{Secure coding guidelines disregard list}
    \label{sec_guidlines}
\small
\begin{tabular}{|l|l|l|l|l|l|}
\hline
    {\textbf{Challenge}}
    & { \textbf{Rule}}
    & { \textbf{Severity}}
    & {\textbf{Likelihood}}
    & { \textbf{Description}}
    & { \textbf{Line number}}
\\ \hline \hline
    \multirow{13}{*}{{ \begin{tabular}[c]{@{}l@{}}Complex \\ Factory\end{tabular}}}
    & { MEM31-C \cite{SEI_CERT}}
    & { Medium}
    & { Probable}
    & {\begin{tabular}[c]{@{}l@{}}Free dynamically allocated memory\\when no longer needed\end{tabular} }
    & { No destructor}
\\ \cline{2-6}
    {} &
    { EXP35-CPP \cite{SEI_CERT}}
    & { High}
    & { Probable}
    & {Do not read uninitialized memory}
    & { 25}
\\  \cline{2-6}
    { }
    & { EXP45-CPP \cite{SEI_CERT}}
    & { High}
    & { Probable}
    & {\begin{tabular}[c]{@{}l@{}}Do not access an object outside  of its \\ lifetime\end{tabular}}
    & { 18}
\\  \cline{2-6}
    { }
    & { MEM51-CPP \cite{SEI_CERT}}
    & { High}
    & { Likely}
    & {\begin{tabular}[c]{@{}l@{}}Properly deallocate dynamically allocated\\ resources\end{tabular}}
    & { 33}
\\ \cline{2-6}
    { }
    & { CTR50-CPP \cite{SEI_CERT}}
    & { High}
    & { Likely}
    & {\begin{tabular}[c]{@{}l@{}}Guarantee that container indices and \\ iterators are within the valid range\end{tabular}}
    & { 18, 25}
\\  \cline{2-6}
    { }
    & { ARR31-C \cite{SEI_CERT}}
    & { High}
    & { Probable}
    & { \begin{tabular}[c]{@{}l@{}}Ensure size arguments for variable \\length length arrays are in a valid range\end{tabular}}
    & { 6}
\\  \cline{2-6}
    { }
    & { CWE-315 \cite{MITRE}}
    & { Medium}
    & { Likely}
    & { Double free}
    & { 33}
\\  \cline{2-6}
    { }
    & { CWE-416\cite{MITRE}}
    & { High}
    & { Likely}
    & { Use after free}
    & { 18, 25}
\\ \hline
    {Sorting}
    & { CWE-208 \cite{MITRE}}
    & { High}
    & { Likely}
    & {\begin{tabular}[c]{@{}l@{}}Observable Timing Discrepancy\end{tabular}}
    & { 7-12}
\\ \hline
    {TOC-TOU}
    & { CWE-367 \cite{MITRE}}
    & { High}
    & { Probable }
    & {\begin{tabular}[c]{@{}l@{}}Time-of-check Time-of-use (TOCTOU)\\ Race Condition\end{tabular}}
    & { 7-12}
\\  \hline
\end{tabular}
  
\end{table*}

\subsection{Sifu Platform}
\label{sec:sifu_platform}

Sifu is a web-based CyberSecurity Awareness Platform \cite{Gasiba2021s}.
This platform was developed in the industry with the aim to raise software developers' awareness on secure coding.
In the industry, this platform is typically embedded in a serious game called CyberSecurity Challenges.
The platform contains several exercises (challenges) that are presented to participants in the form of a project, e.g. in C/C++.
These challenges contain one or more vulnerabilities in the code.
The task of the player is rewrite the code such that it does not contain the vulnerability, while still performing the intended functionality.
All interactions between the player and the platform takes place through the web interface.

Once the player has made changes to the code, he or she can submit the code to the backend, which will analyse the submitted codeand provide feedback to the player.
The goal of the analysis performed in the backend is to assess the presence of cybersecurity vulnerabilities in the code submitted by the player in terms of secure coding guidelines.
The goal of the feedback is, depending on the results of the cybersecurity assessment, to either indicate to the player that the challenge has been solved, or to guide the player to the correct solution by means of hints.

\begin{figure}[http]
    \centering
    \includegraphics[width=1.0\columnwidth]{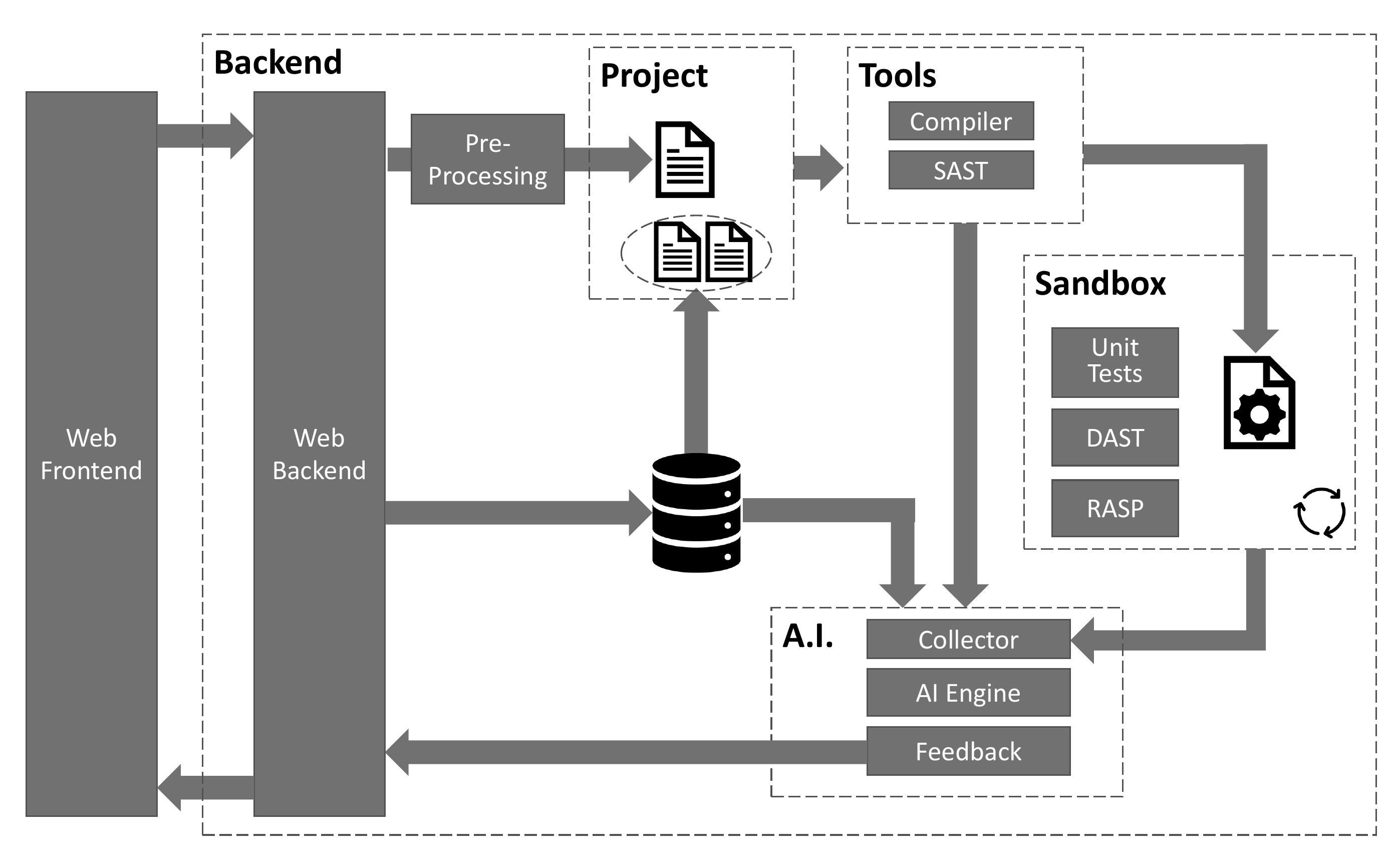}
    \caption{Architecture of the Sifu Platform}
    \label{fig:Sifu_structure}
\end{figure}

\autoref{fig:Sifu_structure} shows the main components of the Sifu platform's backend.
The automatic assessment of challenges is performed using several components: pre-processor, compilers, static and dynamic application security tools, unit tests and run-time application security tests.
An artificial intelligence component collects the results of these tools, performs the cybersecurity assessment and generates hints.
The Sifu platform can be deployed in a local intranet server or in a cloud environment, enabling remote awareness workshops possible.
For further information and details on the platform, we refer the reader to \cite{Gasiba2020s}.

\subsection{Sorting - Time Side Channel}
\label{sec:time_side}
The goal of this challenge is to raise awareness on the player on a well-known security vulnerability called the {\it time side channel} (TSC) \cite{time_side_1}.
A time side-channel occurs when the execution time of an algorithm is different for different inputs of the same size.
In general, side-channels, such as time, can leak data and cause security problems \cite{time_side}.
This type of security vulnerability is critical in embedded systems and programming, since execution time is closely related with power consumption.
Note that time side-channels can be easily observed and measured directly on hardware, such as in an embedded system through an oscilloscope, and measuring power-consumption profiles.
This type of vulnerability can originate in both a poor implementation of hardware components, or non-constant time algorithms implemented in software.
A typical example of a software algorithm that is vulnerable to time side-channels, is a string comparison function which returns, i.e. stops comparing the two strings, whenever the first difference is found.
The run time of the algorithm thus depends on the initial number equal characters.
However, any algorithm that does not run in constant time can leak information.

\begin{lstlisting} [caption={sort.cpp},language=C++, label={sort_code},xleftmargin=1.7\parindent,captionpos=t]
#include <vector>
using namespace std;
 
// This function sorts a vector of int
// Goal: implement the function
void sort(vector<int> &list) {
  size_t i, j;
  for (i = 0; i < list.size(); i++){
    for (j = 0; j < list.size()-1; j++){
      // ...
    }
  }
}
\end{lstlisting}

In this work, we focus on a different algorithm - a array sorting algorithm.
The secure coding guideline which addresses this vulnerability is given by CWE-208: "Observable Timing Discrepancy" \cite{MITRE}.

The source code of the challenge that is presented to the participant is shown in \autoref{sort_code}.
To solve the challenge, the player needs to implement a constant-time sorting algorithm, i.e. a sorting algorithm where the execution time only depends on the number of elements to be sorted and not on their individual values or positions.

\begin{figure}[http]
    \centering
    \includegraphics[width=.95\columnwidth, height=4cm]{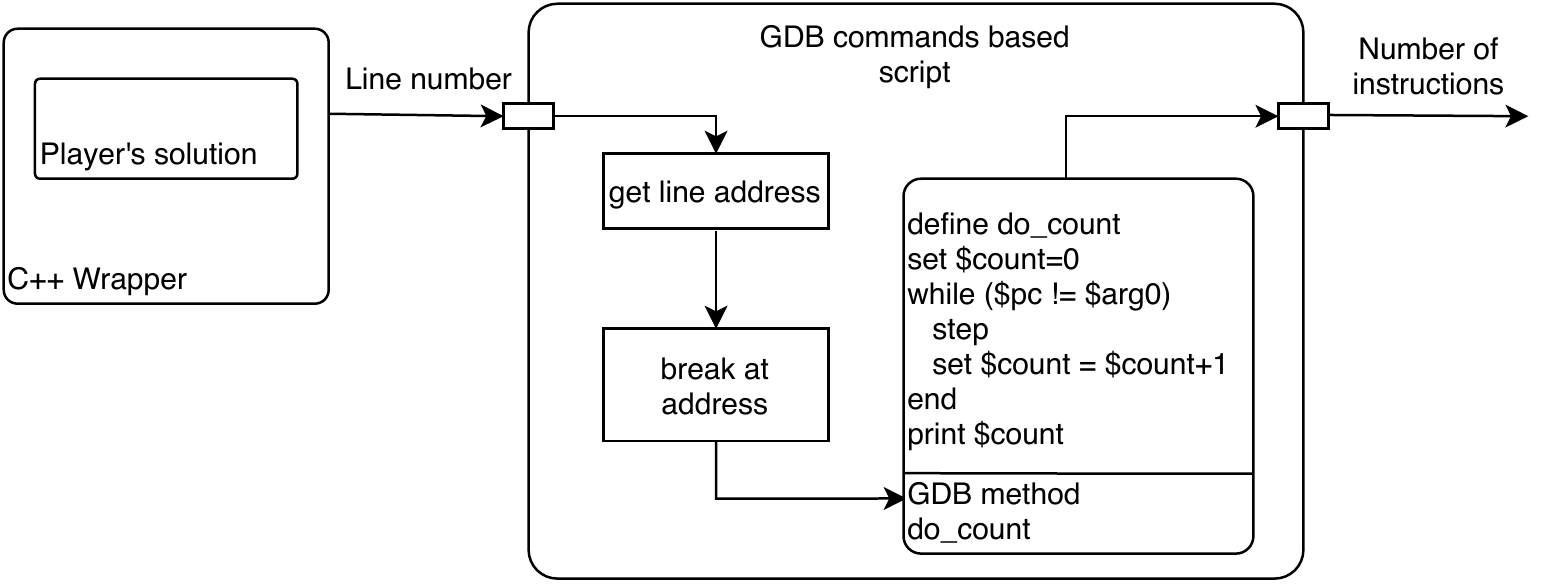}
    \caption{Count instructions diagram}
    \label{fig:count_diagram}
\end{figure}

To address the assessment of the existence of a timing side-channel vulnerability, in code submitted by a player in the Sifu platform, we have searched for existing tools and libraries that could assist in this task.
Two requirements that this tool or library must follow is that 1) it should be independent of an embedded system, and 2) it should be easily run in any environment, such as in the cloud.
A possible solution to this problem is to perform the evaluation by means of an embedded system simulator.
This, however, is not practical since the assessment could incur a considerable delay, making a practical usage in the Sifu platform difficult.
Furthermore, we could not find any library that could easily be used in practice to detect this vulnerability.
Therefore, we propose to use the GNU Debugger, by means of a dedicated python program, as shown in \autoref{fig:count_diagram}.
In this solution, GDB is used to count the number of steps that the entire sorting function needs to execute, i.e. from start to end.
We assume that number of steps required to run the code is highly correlated with the time it takes to execute it, and ignore any possible issues relative to cache misses.
One of the main advantages of this solution is that the measurement is no longer dependent on actual CPU run time, and can be deployed in any environment.
While this method possibly takes more time to perform the assessment than running the code in bare-metal, it will take considerable less time than running an entire embedded hardware simulator.
However, the true result might be dependent on the number of clock-cycles of each assembly instruction.

\autoref{fig:count_diagram} shows an overview on how to perform the cybersecurity assessment of the participant's code.
The code submitted by the participant is embedded in a C/C++ project which contains a wrapper that calls the sorting function.
This small project is then compiled.
A python script starts a GDB session which is used to debug the compiled project.
This script interacts with GDB in the following three steps: 1) create a breakpoint at function call, 2) execute the function step-wise, and 3) step the code until it returns from the function.
During the function's step-wise execution, the total number of steps (iterations) is recorded in an internal GDB variable.
To assess the TSC, the C++ wrapper calls the sorting function with at least two different input vectors of the same size, e.g., using a sorted array and an unsorted array.
The python script measures the number of steps that each function call takes to sort the vector.
Comparing the number of steps in both cases is used to assess the presence of a time side-channel vulnerability.
In our experiments, we have performed the step-wise measurements using the GDB {\it step} and {\it stepi} commands.
The step command executes each line of code, while the  {\it stepi} executes each assembly instruction.

\subsection{Complex Factory}
The main goal of this challenge is to raise the awareness on invalid memory access.
However, this challenge contains several additional vulnerabilities, as shown in table \autoref{sec_guidlines}.
The secure coding guidelines and vulnerabilities were chosen base on their likelihood \cite{SEI_CERT} and adaptability to the challenge.

\autoref{factory_cpp} shows the code that is presented to the player.
The code implements a C++ class that stores complex numbers in an internal buffer of a given maximum size.
The maximum size of the buffer it set at construction time, when an instance of the class is created.
\autoref{sec_guidlines} also details the line numbers where the code's vulnerabilities are present, together with the corresponding SEI-CERT secure coding guidelines \cite{SEI_CERT} and vulnerabilities as defined by the Common Weakness Enumeration (CWE) \cite{MITRE}.

\begin{lstlisting}[caption={ComplexFactory.cpp},label={factory_cpp},language=C++,xleftmargin=1.7\parindent]
#include "FCplx.h"
using namespace std;

/* Constructor allocates a
   container with MAX elements */
FCplx::FCplx(int _max): max(_max)
{
    pos = 0;
    container = new complex<int>[max];
}

/* Stores a complex number in the
   container and returns a reference
   to it */
complex<int>& FCplx::create(int x, int y)
{
  complex<int> a = complex<int>(x,y);
  container[pos++] = a;
  return a;
}
/* Returns a reference to an element 
   stored in the container 
   index 1 returns first element */
complex<int>& FCplx::get(int index){
  return container[index - 1];
}

/* Frees the allocated array. After
   calling this method no further method
   calls are be allowed */
void FCplx::empty()
{
  delete container;
}
\end{lstlisting}

To detect code that disregards the secure coding guidelines presented in \autoref{sec_guidlines}, two distinct methods are used: GCC sanitize flags and security tests.
Sanitize flags are part of dynamic application security testing, and are provided to the compiler at compile time.
They instruct the compiler to generate extra checks that are performed during code execution.
The following are the sanitize flags which we use: {\it address}, {\it leak}, and {\it undefined behavior}.
Address flag adds extra run-time checks on memory addressing, leak flag adds extra run-time checks related to memory allocation, and undefined behavior flag adds extra run-time checks on undefined behavior, as defined by the C++ standard.

The second method used to detect code vulnerabilities is through security testing.
Security tests are used to test specific corner cases, and they try to expose a vulnerability during run-time.
One example of a security test, e.g. to test CWE-315 (double free), is to delete the class variable twice.
A secure solution should catch this problem and react accordingly; however, a poorly implemented solution will cause a double free error, which in turn will trigger the leak sanitizer.
For each secure coding guideline, one or more security test is implemented which tries to trigger the corresponding vulnerability.

\subsection{Race Condition}
The goal of this challenge is to raise the awareness on race condition vulnerabilities.
A race condition can occur when two or more concurrent processes try to access a shared resource, whereby at least one process tries to modify the shared resource.
The time between the resource is read and the resource is modified is known as race window or critical section.
Consequences of exploiting this types of vulnerability include denial-of-service, and privelege escalation.

While race conditions typically occur in shared memory, we propose a challenge based on files.
In this case the shared resource is a file and the vulnerability is called a time-of-check, time-of-use (TOCTOUC).
The security vulnerability is listed in the CWE database under CWE-367 \cite{MITRE}.
\autoref{toctou} shows the challenge that is presented to the player.
To solve the challenge, the player needs to perform two steps: check if a file exists and, if it exists, modify its permissions.
The race window occurs between the time that the existence of the file is checked until the attributes are modified.
The problem is that a malicious user can change the file between the two operations and, therefore the code changes the permissions of the wrong file.
Previous research that addresses this vulnerability on real-time embedded systems is \cite{embedded_1, embedded_2}.
Note: one possible solution to this challenge is by using the fchmod, instead of the chmod function.

\begin{lstlisting} [caption={set\_permissions.cpp},language=C++, label={toctou},xleftmargin=1.7\parindent,captionpos=t]
/* Check if the file exists, and change
   the mode of the file. Return true if
   everything was successful */
bool setPerm(char *fName, mode_t mode){
  // Check if the file exists
  FILE *f_ptr;
  // Change the mode
  if (chmod(fName, mode) == -1) {
    // Handle error ...
    return false;
  }
  return true;
}
\end{lstlisting}

\begin{figure}[http]
    \centering
    \includegraphics[width=1.0\columnwidth]{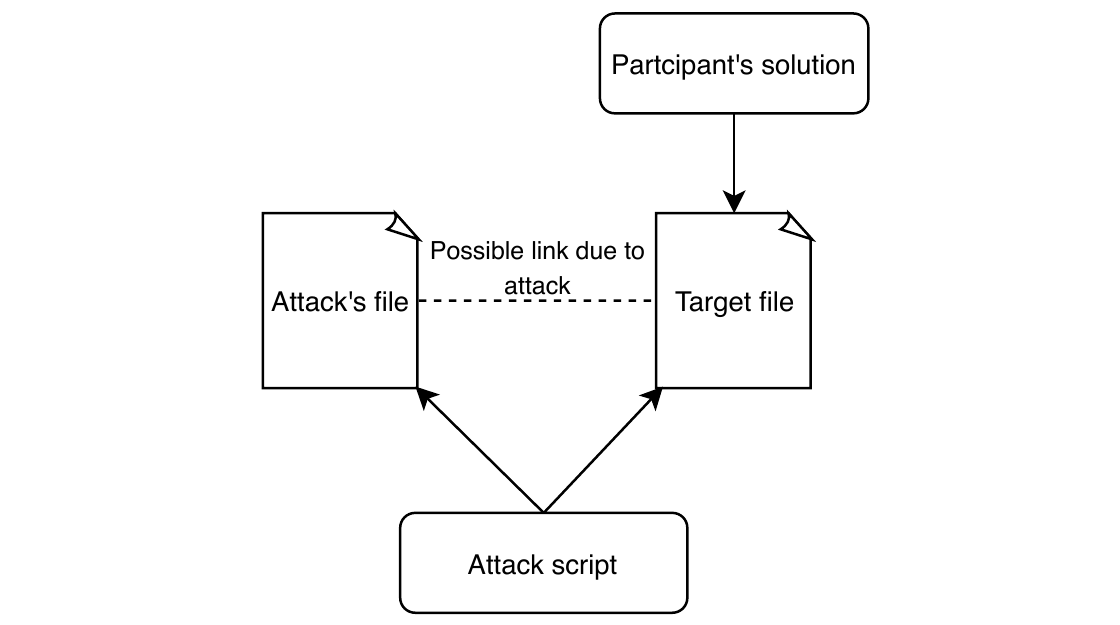}
    \caption{Top-level challenge structure}
    \label{fig:toc_struct}
\end{figure}

This type of vulnerability can potentially be detected by means of static application security testing tools.
However, to the best of our knowledge, except for commercial solutions, there is no open-source tool that can detect this type of vulnerability \cite{moerman2018evaluating}.
Furthermore, static application security testing tools are known to be unreliable \cite{oyetoyan2018myths,moerman2018evaluating}.
In light of this, we propose a simple solution based on an attack script and a wrapper function, as shown in \autoref{fig:toc_struct}.
The wrapper function calls the participants' solution multiple times while, in parallel, an attacker script is running and swapping the two files in an endeless loop.
Both the wrapper and the attacker script stop executing when one of the following conditions is met: (1) the permissions are changed in the wrong file, or (2) a maximum number of iterations is achieved.
Condition two will occur if there is no vulnerability in the code of the player, or the race condition could not be achieved, i.e. the vulnerability was not detected after all the iterations.

Note that, since the race condition is not reliably triggered, several tries need to be performed.
We expect that the more iterations are performed, the higher the probability to detect the the vulnerability.
In \autoref{sec:evaluation} we present the results of evaluation on the trade-off between the detection probability and the total number of iterations.

Another possible solution to the detection problem is by artificially modifying the player's code and injecting a delay in the code's critical section.
Although this solution can potentially increase the reliability to detect the vulnerability, it requires, however, a modification of the source code submitted by the participant.
Furthermore, the method and implementation details to achieve this reliably is currently an open topic.

\section{Evaluation}
\label{sec:evaluation}
Validation of the presented cybersecurity assessment methods was performed through computer experimentation.
Additionally, the challenges were deployed in an academic environment, followed by semi-structured interviews and a survey.
In this section we give details on the evaluation process.

\subsection{Evaluation of Design}
For the Time Side-Channel challenge, an evaluation took place by means of two possible implementations of the sorting algorithm: with and without a time side channel.
The algorithm containing the vulnerability was a standard bubble-sort algorithm, while the implemented solution was a bubble sort algorithm modified to swap elements with the same index, thus increasing run-time and avoiding the time side-channel vulnerability.
The input to the sorting algorithm consisted of three random permutations of an integer vector size of 5 elements.
The process was repeated 1000 times for both algorithms and for the \textit{step} and \textit{stepi} GDB commands respectively.
Furthermore, the total execution required to get an answer from the backend was measured.

For the Race Condition challenge, we focus on the evaluation of the detection probability of the vulnerability.
To compute this probability, we ran the python script 1000 times, and recorded the number of iterations required to detect the vulnerability.
Based on these results, we can compute the cumulative density function of the detection probability $c(n)$ by normalizing the number of times that the vulnerability was detected in less than {\it n} cycles.

Since the Complex Factory challenge uses standard detection mechanisms, we have not performed an evaluation step for the security assessment.

All the tests were conducted on a PC with the following specifications: Ubuntu 18.04.4 LTS on an Intel i5-3427U CPU running at 1.80GHz with four cores and 8 GB of RAM.

\subsection{Challenge Deployment in Academia}

We have deployed the challenges in the Sifu Platform and asked several students, without previous industry experience, to evaluate them during two runs.
\autoref{tab:part_into} shows details on the demographics of these runs.
\begin{table*}[hb]
\small
\caption{Participants' information}
\label{tab:part_into}
\resizebox{\textwidth}{!}{
\begin{tabular}{|l|l|l|l|l|l|l|l|}
\hline

\textbf{No.} & \textbf{Start Date} & \textbf{End Date} & \textbf{Participants} & \textbf{Where} & \textbf{Age Range} & \textbf{Field of Study} & \textbf{Educational Level} \\ \hline \hline
1 & 5 Nov 2020 & 10 Nov 2020 & 12: Germany & Online & 20-28 & \begin{tabular}[c]{@{}l@{}}Electrical engineering,\\ Computer engineering,\\ Informatics\end{tabular} & \begin{tabular}[c]{@{}l@{}}Bachelor's degree\\ Master's degree\end{tabular} \\ \hline
2 & 16 Nov 2020 & 23 Nov 2020 & 4: Germany & Online & 22-25 & \begin{tabular}[c]{@{}l@{}}Computer engineering,\\ Communications engineering\end{tabular} & Master's degree \\ \hline
\end{tabular}
}
\end{table*}

The participants were first given a short introduction to the platform, and were given instructions on how to use it.
Next, the participants were given time to check and familiarize with the platform, and to solve the challenges.
After solving the challenges, a semi-structure interview took place.
Participation in the semi-structured interview was not mandatory, and the collected answers were anonymized.

The participants were also asked to answer a small survey consisting of eleven questions.
The questions that were asked in the survey are shown in \autoref{tab:final_questions}.
Survey feedback answers were gathered using Google Forms.
The answers to the questions were based a 5 point Likert scale \cite{likert} for agreement, i.e. {\it strongly disagree}, {\it disagree}, {\it neutral}, {\it agree}, and {\it strongly agree}.
The questions are adopted from \cite{Gasiba2021s} and extended to target the academia.
The adoption of the same questions allows, in the results section, to compare the answers from academia to answers from the industry.

\begin{table}[http]
\small
\caption{Survey questionnaire}
\label{tab:final_questions}
\resizebox{\columnwidth}{!}{
\begin{tabular}{|l|l|}
\hline
\textbf{} & \textbf{Survey Question}                                                                                             \\ \hline
\hline
Q1         & \begin{tabular}[c]{@{}l@{}}Paying attention to secure coding increases my code \\ quality\end{tabular}               \\ \hline
Q2         & University teaching includes awareness in secure coding                                                              \\ \hline
Q3         & \begin{tabular}[c]{@{}l@{}}I learned new techniques and principles of secure \\ software development\end{tabular}    \\ \hline
Q4         & \begin{tabular}[c]{@{}l@{}}I know how to use the information about secure coding \\ guidelines\end{tabular}          \\ \hline
Q5         & I understand the importance of secure coding guidelines                                                              \\ \hline
Q6         & \begin{tabular}[c]{@{}l@{}}Focusing on the challenges improves my practical secure\\ coding skills\end{tabular}      \\ \hline
Q7         & \begin{tabular}[c]{@{}l@{}}I have learned about new issues that I would like to \\ check in my own code\end{tabular} \\ \hline
Q8         & \begin{tabular}[c]{@{}l@{}}I know where I can find more information about secure\\ coding guidelines\end{tabular}    \\ \hline
Q9         & \begin{tabular}[c]{@{}l@{}}The learning goals of the challenges were clearly\\ explained\end{tabular}                \\ \hline
Q10        & The help from the virtual coach was adequate                                                                         \\ \hline
\end{tabular}
}
\end{table}

In total, six participants responded to the planned semi-structured interview.
The semi-structured interview questions were based on the following questions: {\it what is the most significant advantage in participating in these challenges}, {\it  what did not go well and you would like to change}, and {\it do you think that secure coding awareness increases the code quality overall}.

\section{Results}
\label{sec:four}

This section show the results on the evaluation of the proposed vulnerability assessment schemes for different the challenges.
This section also shows the analysis of the survey questions, and result from the semi-structured interviews.
We also present a comparison of our survey results with two similar surveys by Gasiba et al. which were held in an industrial context.

\subsection{Sorting - Time Side Channel}
\label{sec:sort}

\begin{figure}[http]
    \centering
    \includegraphics[width=1.0\columnwidth]{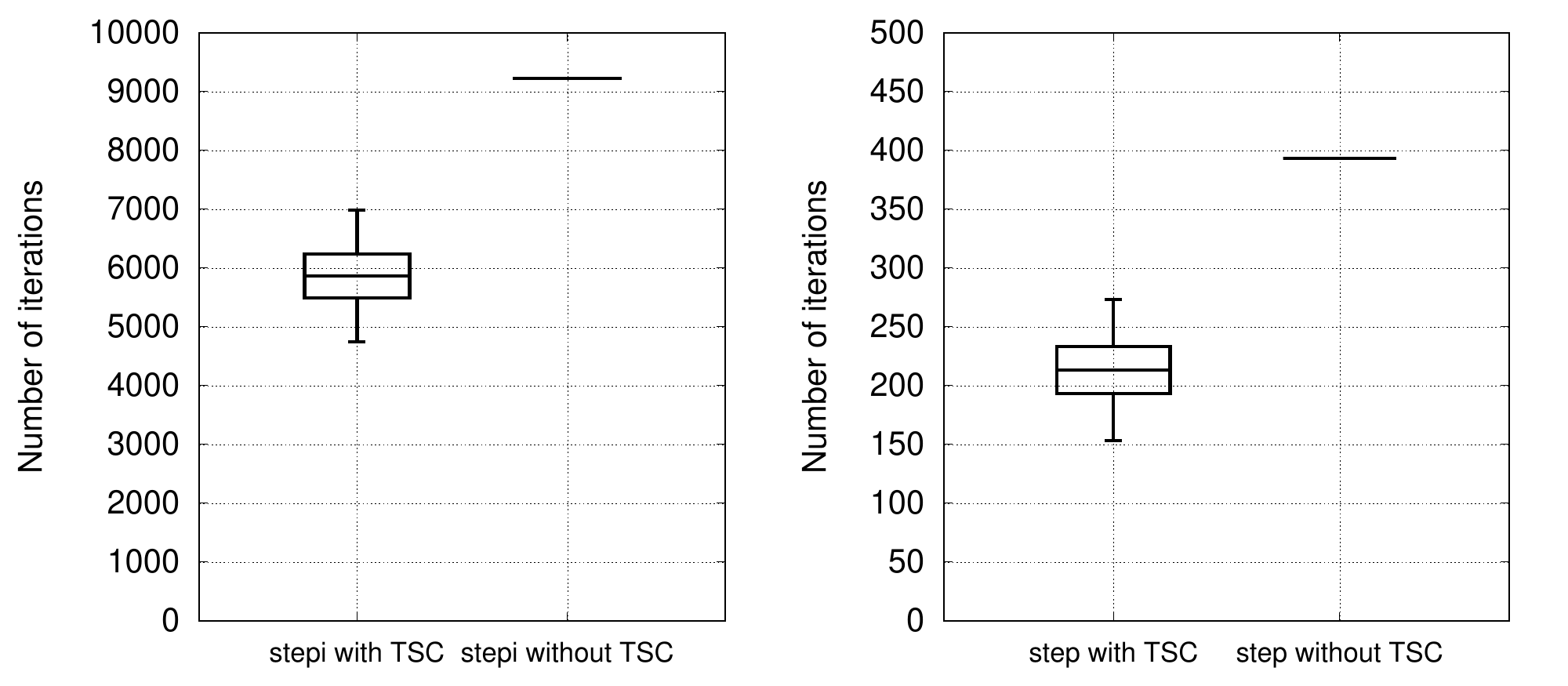}
    \caption{Step instruction comparison}
    \label{fig:instruction_comparison}
\end{figure}

\autoref{fig:instruction_comparison} presents the results when using \textit{step} (right plot) and \textit{stepi} (left plot) GDB commands respectively.
Both instructions, step and stepi, show only a line (i.e. single value) when there is no time side-channel vulnerability present in the code.
This result comes as a consequence of having the same number of instructions for every input.
Furthermore, the vertical lines show the worst and best case for no TSC.
The observable difference in instruction count comes as a consequence that the stepi instruction needs to execute every assembly instruction of the source code.

The stepi GDB command takes more iterations than the step GDB command to perform the assessment of the vulnerability.
This result is expected, since the stepi command executes the code one assembly instruction at a time, while the step command executes the code one source code line at a time. 
Due to their nature, the stepi command is more precise than the step command; however this represents a trade-off between precision and speed, since the stepi command is more than twenty two times slower than the step command.

\autoref{fig:instruction_comparison} also shows that, for the stepi GDB command, an increase in the number iterations between 31.8\% and 95.3\% between code without TSC and code with TSC is observed.
The same figure shows that for the step command, the increase in the number of iterations is between 44.4\% and 157\% for code without TSC and code with TSC.

The second aspect to check the design is the reasonable delay after the participant has submitted a solution.
We have measured the execution time of this implementation when running it in the Sifu platform. \autoref{tab:execution_time} shows the captured measurements together with 1st and 3rd quartile. Same cases were covered as in the \autoref{fig:instruction_comparison}. 

\begin{table}[http]
\caption{Execution time measurement}
\label{tab:execution_time}
\small
\begin{tabular}{|l|l|l|l|}
\hline
\textbf{Test case}              & \textbf{Mean exec time {[}s{]}} & \textbf{Q1}& \textbf{Q3}\\ \hline \hline
Step with TSC     & 1.81 &    1.76 & 1.86                     \\ \hline
Step without TSC  & 2.19 &     2.18 & 2.20                   \\ \hline
Stepi with TSC   & 3.74 &      3.44 & 3.91                \\ \hline
Stepi without TSC & 5.43 &       5.39 & 5.48              \\ \hline
\end{tabular}

\end{table}

\subsection{Race Conditions}
\begin{figure}[http]
    \centering
    \includegraphics[width=1.0\columnwidth]{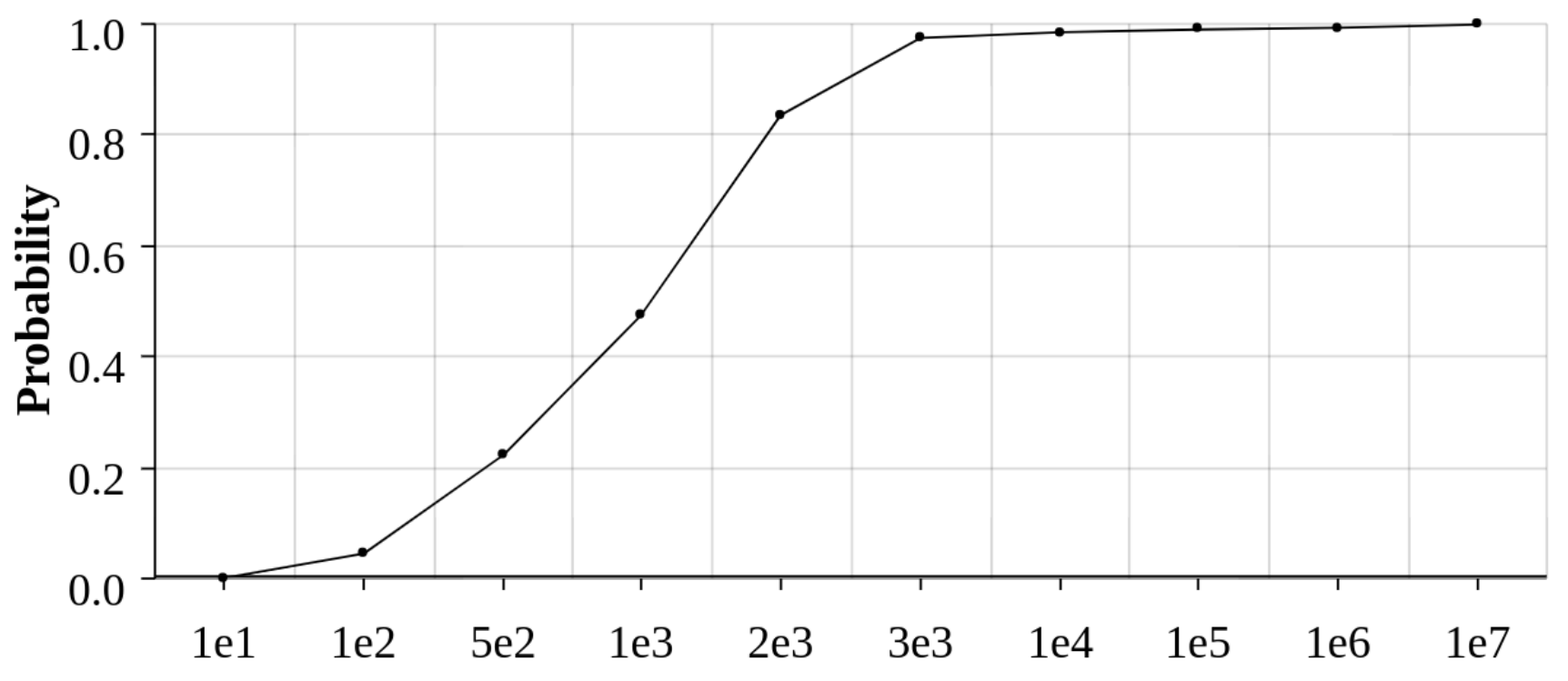}
    \caption{Success probability related to number of tries}
    \label{fig:race}
\end{figure}

In the TOC-TOU Race Condition Challenge, we evaluated the detection probability of the attacking script.
The results are shown in \autoref{fig:race}.
The x-axis of this figure shows the number of tried, i.e. iterations, performed to detect the vulnerability, while the y-axis shows the probability of detection.

This graph shows that, to achieve a detection rate of 99\%, more than 3000 iterations are required.
Since the execution time for a single round of this assessment method is very small (in the order of microseconds), we recommend to implement more than 10,000 iterations in a practical scenario.
Our results show that, this value is reasonable and does not lead to considerable delay in the backend.

\autoref{fig:race} can be used as a guideline by practitioners who wish to implement this detection mechanism in their own deployments.
Since the actual running time depends on the speed and on the CPU where the test is run, we recommend that practitioners start with the value 10,000 and run their own evaluation of the number of iterations that are required to achieve a detection probability of 99\% or higher.

\label{sec:race_res}

\subsection{Survey Results}
\begin{figure*}[ht]
    \centering
    \includegraphics[width=2.0\columnwidth]{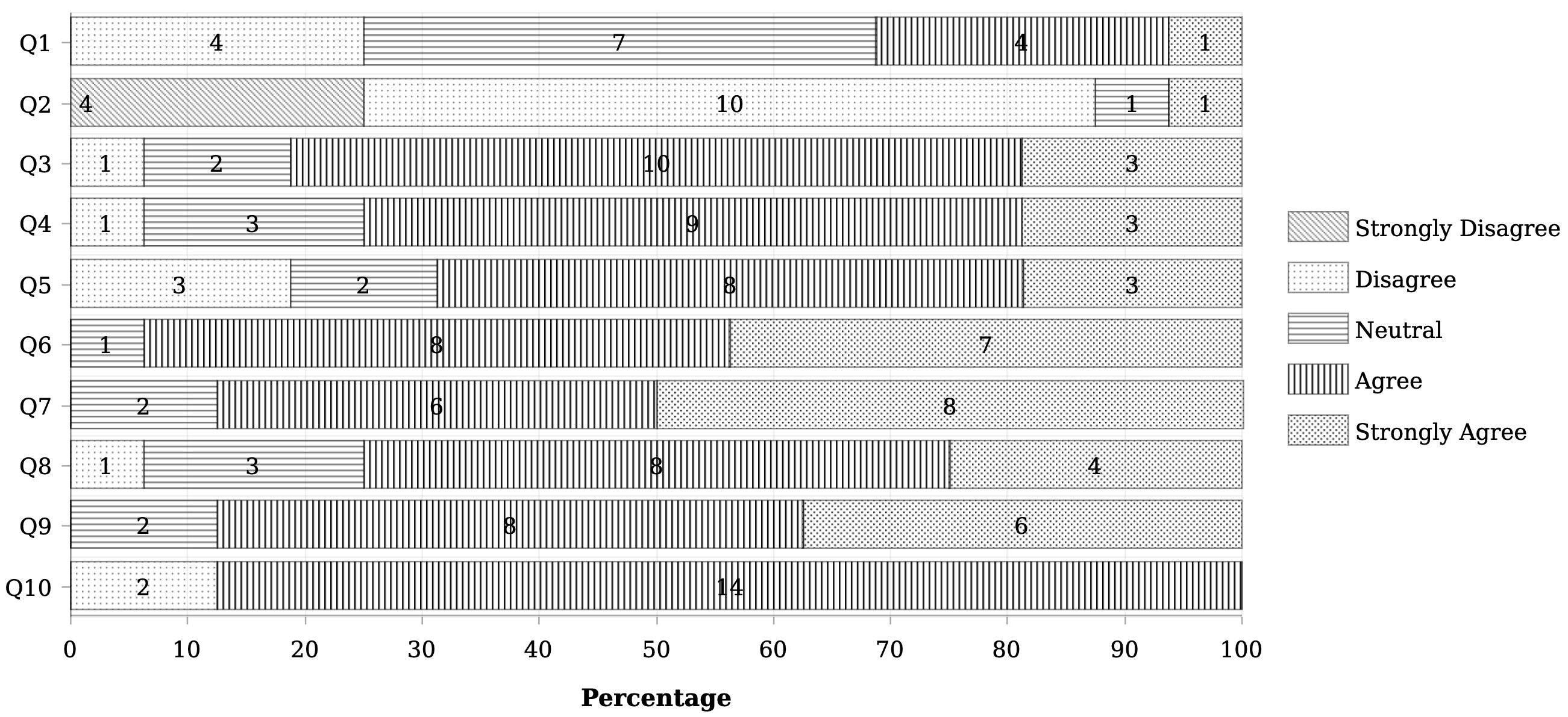}
    \caption{Survey Results}
    \label{fig:survey_1_results}
\end{figure*}

\autoref{fig:survey_1_results} shows the results of the survey which was based on the questions introduced in \autoref{tab:final_questions}.
Our preliminary results show, except for Q1 and Q2, an overall agreement with the survey question.
The questions that include the highest agreement are Q6, Q7, Q9, and Q10.
This means that the participants agree that the assistance provided by the virtual coach is adequate, thus providing a good indicator of the suitability of the proposed vulnerability detection methods.
Participants also agree that focusing on secure coding challenges improves their security coding skills, therefore also their awareness on the secure coding guidelines.
Finally, the participants have learned new vulnerabilities that they would like to check in their own code.
This gives a good indicator that these types of games can motivate a positive behaviour towards secure coding.

The aspects there received the lowest amount of agreement, though still overall positive, are related with Q4 and Q8.
This is not surprising since the Sifu platform is developed for the industry, and motivates the participants to find additional information on secure coding using internal resources; the same cannot be said for the academia, where further information should be searched in online forums or scientific publications.
It is however surprising that the participants claim that playing the challenges allows them to know where to find more information about secure coding, and how to use this information.

We also observe that, in general, there is still a high level of neutral answers.
The reason for this is not well understood and further research is necessary to understand this issue.
Also, for Q1, it is surprising that the participants are not entirely sure if paying attention to secure coding increases their code quality.
Although one possible reason for this might be the lack of specific training in secure coding and code quality, this aspect needs more search.
Finally, another surprising result is the one obtained in Q2, where the participants claim that university teaching does not include awareness on secure coding.
This result is an indicator that this awareness campaigns are necessary to be held in the industry, and a specialized course that covers secure coding and secure coding guidelines might be beneficial.
Attending this course would better prepare the students for a future career in the industry.

\subsection{Comparison to Previous Work}
\autoref{fig:comparison} shows a comparison of the current work, and the work by Gasiba et al. (see \cite{Gasiba2020e}, and \cite{Gasiba2020s}).
The main difference is that, in the current work, we evaluate the CyberSecurity Challenges in the academia, while previous work evaluates these in an industrial context.
Both the current work as the previous work show a general agreement in relation to all the survey questions.
However, one important difference is the fact that Q5 has a higher agreement level for the industry as in the academia.
We think that the main reason for this difference might be related with the fact that the usage of secure coding guidelines are typically mandated by security policies in the industry, e.g. as a result of requirements from cybersecurity standards, while this is not the case in academia.
We also observe that Q4 and Q8 exhibit a large amount of neutral answers for both academia and also for the industry.
The reason behind this is not entirely understood and requires further investigation.

In terms of disagreement, we observe that Q3, Q7, and Q8 have a similar amount of disagreement in both the academic setting and also in the industrial setting.
One common positive point in all the surveys, industry and academia, is the fact that the support by the coach is seen as an important factor.
The factor that has the largest amount of difference between the academia and industry is Q8 - knowing where to find more information;
Based on our experience, we think that this might be related with the fact that in an industrial environment, there are internal procedures related with secure software development life-cycle which establish where further information can be obtained.
This might justify the higher value for academia and lower value for the industry.

Surprisingly there is about 8\% of negative answers in relation to Q9 - the goals of the challenge are clearly stated - for the industry, while 0\% for academia.
This effect might be related with the diversity of software developers in the industry; however, further investigation is needed to understand this discrepancy.

In summary, although there are some differences observed in the results of academia vs industry, overall there is a large agreement for all the survey questions.
This fact gives a good indication that the Sifu platform and its challenges are well suited for both academia and industry.

\begin{table*}[http]
\centering
\caption{Comparison with previous work}
    \label{fig:comparison}
\resizebox{.95\textwidth}{!}{
\begin{tabular}{|c|c|c|c|c|c|c|c|c|c|c|c|} 
\hline
\multicolumn{4}{|c|}{\textbf{Present work}}                                              & \multicolumn{4}{c|}{\textbf{Gasiba et al. \cite{Gasiba2020e}}}                                        & \multicolumn{4}{c|}{\textbf{Gasiba et al. \cite{Gasiba2020s}}}                                         \\ 
\hline
\textbf{Question} & \textbf{Negative} & \textbf{Neutral} & \textbf{Positive} & \textbf{Question} & \textbf{Negative} & \textbf{Neutral} & \textbf{Positive} & \textbf{Question} & \textbf{Negative} & \textbf{Neutral} & \textbf{Positive}  \\ 
\hline
\hline
Q1                & 25.0\%            & 43.7\%           & 31.3\%            &  \cellcolor{light-gray}                 &  \cellcolor{light-gray}      &  \cellcolor{light-gray} & \cellcolor{light-gray}    & \cellcolor{light-gray}    &  \cellcolor{light-gray}    &  \cellcolor{light-gray}  & \cellcolor{light-gray}     \\ 
\hline
Q2                & 87.6\%            & 6.2\%            & 6.2\%             &  \cellcolor{light-gray}                 &  \cellcolor{light-gray}      &  \cellcolor{light-gray} & \cellcolor{light-gray}    & \cellcolor{light-gray}    &  \cellcolor{light-gray}    &  \cellcolor{light-gray}  & \cellcolor{light-gray}     \\
\hline
Q3                & 6.2\%             & 12.5\%           & 81.3\%            & Q1.1              & 12.5\%            & 7.1\%            & 80.4\%            & X1                & 0.0\%             & 10.0\%           & 90.0\%             \\ 
\hline
Q4                & 6.2\%             & 18.8\%           & 75.0\%            & Q6.1              & 8.9\%             & 28.6\%           & 80.3\%            &    \cellcolor{light-gray}    &       \cellcolor{light-gray}  &   \cellcolor{light-gray}  &    \cellcolor{light-gray}  \\ 
\hline
Q5                & 18.7\%            & 12.5\%           & 68.8\%            & Q10.9             & 0.0\%             & 5.3\%            & 94.7\%            & X9                & 0.0\%             & 0.0\%            & 100.0\%            \\ 
\hline
Q6                & 0.0\%             & 6.2\%            & 93.8\%            & Q7.1              & 3.6\%             & 14.3\%           & 82.1\%            & F2                & 0.0\%             & 4.0\%            & 96.0\%             \\ 
\hline
Q7                & 0.0\%             & 12.5\%           & 87.5\%            & Q8.1              & 10.7\%            & 5.4\%            & 83.9\%            &   \cellcolor{light-gray}  &   \cellcolor{light-gray} &  \cellcolor{light-gray} &  \cellcolor{light-gray} \\ 
\hline
Q8                & 6.3\%             & 18.7\%           & 75.0\%            & Q9.1              & 12.5\%            & 31.2\%           & 55.4\%            & X8                & 0.0\%             & 10.0\%           & 90.0\%             \\ 
\hline
Q9                & 0.0\%             & 12.5\%           & 87.5\%            & Q11.1             & 8.9\%             & 8.9\%            & 82.2\%            & F8                & 8.0\%             & 8.0\%            & 84.0\%             \\ 
\hline
Q10               & 12.5\%             & 0.0\%            & 87.5\%           & Q13.1             & 1.8\%             & 12.5\%           & 85.7\%            & X6                & 0.0\%             & 0.0\%            & 100.0\%            \\ 
\hline
\hline
\multicolumn{4}{|c|}{16 participants}                                        & \multicolumn{4}{c|}{56 participants}                                         & \multicolumn{4}{c|}{25 participants}                                          \\
\hline
\end{tabular}
}
\end{table*}

\subsection{Semi-Structured Interviews}
After using the Sifu platform and solving each of the proposed challenges, participants were interviewed.
Analysis of the participants' answers resulted in their classification into three groups: {\it Benefits}, {\it Application in Embedded systems}, and {\it Drawbacks}.
\autoref{tab:quotes} shows the top 10 quotes from participants, along with a mapping to the individual groups.
The group {\it Benefits} covers the user experience and what the participants thought was the most beneficial outcome of the challenges.
Group {\it Application in Embedded systems} was related to our primary goal, which was not explicitly mentioned, with a reason to see if the participants see a possible application of this platform for training secure coding in embedded systems. Finally, the group {\it Drawbacks}, covers negative aspects experienced by the participants.
The feedback that we have collected allows to understand the strong and weak points of the platform, and also to improve the it further.

\begin{table}[http]
\caption{Quotes from Participants}
\label{tab:quotes}
\footnotesize
\resizebox{\columnwidth}{!}{
\begin{tabular}{|m{.5cm}|p{5.4cm}|p{1.6cm}|}
\hline
{ \textbf{No}} & { \textbf{Quote from Participant}} & { \textbf{Group}}       \\ \hline
\hline
{ 1}           & { I believe that accounting for security while coding is beneficial in writing more efficient code.}  & \multirow{8}{*}{ Benefits} \\ \cline{1-2}
{ 2}           & { Learning really useful strategy to make significantly more quality code}   &                                                     \\ \cline{1-2}
{ 3}           & { The best thing is that platform acts as a game} &                                                        \\ \cline{1-2}
{ 4}           & { I think it is good to be exposed to an important aspect of coding which is handling or accounting for security issues or potential vulnerabilities} & \\ \hline
{ 5}           & { Suitable for training secure coding on embedded systems} & \multirow{6}{1.6cm}{ Embedded Programming} \\ \cline{1-2}
{ 6}           & { This could be used as a training platform for some courses at my university} & \\ \cline{1-2}
{ 7}           & { This will be a real advantage for users who are working on security critical systems.} & \\ \hline
{ 8}           & { The hints given by the chat bot were not always accurate or precisely leading to the nature of the problem in the code.} & \multirow{7}{*}{Drawbacks} \\ \cline{1-2}
{ 9}           & { The user interface is minimalistic for nowadays  standards.} &             \\ \cline{1-2}
{ 10}          & { Make a more sophisticated user interface and experience} &                 \\ \hline
\end{tabular}
}
\end{table}

The grouping of certain answers and comments answered how participants perceive the platform and see a possible application 
in embedded programming.
The main outcome of the semi-structured interviews are as follows.
For the perceived benefits of the platform, the fact that it is used as a game was positively perceived. 
Furthermore, playing the games exposes the participants to secure coding tasks, which is also perceived as being beneficial.
In therms of the embedded programming group, some participants suggested that the platform might be used in a standard course at the university, which might be beneficial for those who work on security critical systems, e.g. in critical infrastructures.

Although most of the feedback was positive, we collected some drawbacks.
The two main negative points are related with the user interface, and with the precision of the feedback given by the virtual coach.
Related to quotes 9 and 10, interviewees gave a more detailed answer in our discussion.
One interviewee said that having a standard debugging tool would ease the challenges and increase the learning factor.
As well to compete with other online learning platforms, more details on the design have to be applied. 

The collected positive and optimistic answers indicate that participating in these challenges can potentialy lead to raising secure coding awareness in academia.
The answers are also encouraging towards validating the suitability of the proposed vulnerability assessment methods.

\subsection{Discussions}
In this work we have presented the implementation and evaluation of cybersecurity challenges for the Sifu platform.
The implementation of the challenges covered technical aspects on how to evaluate the presence of a given vulnerability in source code, while the evaluation was performed in several ways: computer experiments, survey, semi-structured interviews and comparison to previous work.

In terms of technical implementation, we have discussed a mechanism to detect time side-channel, and race conditions.
The main idea for the detection of time side-channel is to use a standard debugger, such as GDB, together with a wrapper and a python script to control the debugger.
By stepping through the code, we can evaluate the number of steps that the algorithm would require to run from beginning to the end.
In the results section we present practical advice for practitioners who want to implement this method, in particular we discuss about the trade-off between precision and execution time.
Notice that we propose a simplified method to assess the presence of a time-side channel.
In practice, other effects will impact the running time of the algorithm, such as number of cycles per instruction, hyper-threading, cache misses, and CPU internal pipelines, and therefore the presence of time-side channels.
However, although the proposed method is simple, it allows to raise awareness on time-side channels through a serious game.
Furthermore, the reason why we observe highly correlated results between the "step" and "stepi" method is related to the fact that the implemented algorithm mostly uses mathematical operations and no function calls.

For the race condition vulnerability, we propose to write a script that attacks the function written by the participant.
We discuss the trade-off between probability of detecting the vulnerability and execution time.
The reason why execution time is crucial, is that the Sifu platform is an interactive platform.
The higher the delay in the backend, the less interactive the platform becomes, which could lead to problems with user experience.

One major finding in our survey results is the fact that the students claimed a  lack of awareness on secure coding during courses at the university.
While this result cannot be generalized to every university and every course, it does raise the need to address this topic in general, as the students of today are the workforce of the future.
The survey results also indicate that the Sifu platform might be suitable to address this awareness in secure coding at the university.
In particular, we have received positive indications that the platform could be integrated into the standard teaching curricula.
A comparison with two previous surveys performed in an industrial setting was also discussed.
Both the studies performed in the academia as also in the industry show encouraging results on the suitability of the platform as a means to raise awareness on secure coding.
However, there are small differences between the surveys that indicate small discrepancies.
We believe that a practitioner who wishes to deploy or refine these types of challenges in the academia, can find valuable information in these studies, to guide in their decision making.

\vspace{-.35em}
\subsection{Threats to Validity}
\label{sec:validity}

Possible threats to our conclusions include: number of participants, study field and experience, and participant bias.
Our preliminary results are mostly positive.
This might be related to the relatively low number of participants to the survey, since it was not mandatory.
Therefore, some negative comments might not have been been captured.
Participants have different study fields, backgrounds, and are at different levels in their studies.
Although the number of participants is limited, it is in line with comparable empirical studies.
Therefore, some participants might find the challenges easy, while other might have more difficulties to solve them. 
This might lead to different answers to the survey and also to the semi-structured interview and, therefore.
A more detailed analysis and research on the answers given by the different groups would be required to understand possible bias effects.

Finally, since the participants are aware of the purpose of the study, a positive bias cannot be discarded.
In particular, participants might respond to the survey in a way that they think the authors expect them to answer.
Nevertheless, the results obtained in this work are in alignment with previous work that was done in an industrial environment.
Therefore, we do not think that considerable different conclusions from the ones presented in this work would be obtained by increasing the number of survey participants.

\section{Conclusions}
\label{sec:five}

Security bugs, when exploited, often can lead to serious software vulnerabilities.
Nowadays, secure coding guidelines exist to teach software developers and make them aware of software vulnerabilities and how to write secure code that avoids these vulnerabilities.
However, not all software developers are knowledgeable about these or secure coding standards in general.
This is true for the industry and, the present study also finds out that this might also be true in academia.

To address this issue, this work extends previous research conducted on the Sifu platform.
We introduce three challenges with security vulnerabilities common in embedded programming that can be integrated into university teaching curricula.
This paper consists of two main parts: 1) a brief description of how to implement the challenges in terms of evaluation of the presence of the vulnerability in the participant's source code, and 2) an evaluation of the challenge design and evaluation of the challenges in an academia setting.

In the first part, three different C++ challenges were implemented on Sifu's platform.
The implementation process and decision-making were briefly explained.
A critical part of our work is how to test particular security vulnerabilities that are presented in the challenges.
We give a detailed explanation of the architecture and the implementation of testing particular vulnerabilities.

In the second part, we evaluate the design from a technical view, and an empirical view.
Our results indicate that the proposed methods can be used to detect the challenges' vulnerabilities.
Additionally we give practical advice on the implementation of the challenges.
Through two trial runs in academia, we collected answers on a survey and performed a semi-structured interview.
Our preliminary results give a good indication that the challenges are adequate to raise secure coding awareness for students.
Additionally the results for the semi-structured interview give a positive indication on the suitability of the challenges for academia, and give good insight into future improvements, e.g. on user experience.
Finally, we have performed a comparison with previous studies, thus comparing results from industry to results from academia.
While the majority of the results indicate a good agreement between industry and academia, the small differences between both can serve as a guide to practitioners who wish to deploy the Sifu platform in an academic setting.

As further steps, the authors would like to investigate the usage of popular open-source static code analysis tools, and tapping system calls to improve the platform's detection mechanisms.
Furthermore, the authors would like to address these tools' usage to improve the platform's hint mechanism. 

\begin{acks}
The authors would like to thank the survey participants for their useful and insightful discussions and for their participation in the test run.
This work is partially financed by national funds through FCT - Fundação para a Ciência e Tecnologia, I.P., under the projects FCT UIDB/04466/2020 and UIDP/04466/2020. Furthermore, the fourth author thanks the Instituto Universitário de Lisboa and ISTAR, for their support. 
\end{acks}

\bibliographystyle{ACM-Reference-Format}
\bibliography{000_paper}

\end{document}